\documentclass[11pt]{article}
\usepackage{moriond,epsfig,amsmath}
\def\Journal#1#2#3#4{{#1} {\bf #2}, #3 (#4)}

\def\NPB{{\em Nucl. Phys.} B}
\def\PLB{{\em Phys. Lett.}  B}
\def\PRL{\em Phys. Rev. Lett.}
\def\PRD{{\em Phys. Rev.} D}

\def\be{\begin{equation}}
\def\ee{\end{equation}}
\def\bea{\begin{eqnarray}}
\def\eea{\end{eqnarray}}

\begin{document}
\vspace*{4cm}
\title{MEASUREMENTS OF THE MASSES, LIFETIMES AND MIXINGS OF B HADRONS AT THE TEVATRON}

\author{ S. MALDE on behalf of the CDF and D0 collaborations}

\address{Department of Physics, Denys Wilkinson Building, 1 Keble Road,\\
Oxford OX1 3RH, United Kingdom}

\maketitle\abstracts{
The Tevatron, with $p\overline{p}$ collisions at $\sqrt{s}=$1.96 TeV, can produce all flavors of B hadrons and allows for unprecedented studies in the B physics sector. The CDF and D0 collaborations have more than 5 fb$^{-1}$ of data recorded. I present here a selection of recent results on the masses, lifetimes and mixings of B hadrons using between 1.0 and 2.8 fb$^{-1}$ of data.}



\section{Observation of the $\Omega_b$} 
\begin{figure}
\begin{center}
\psfig{figure=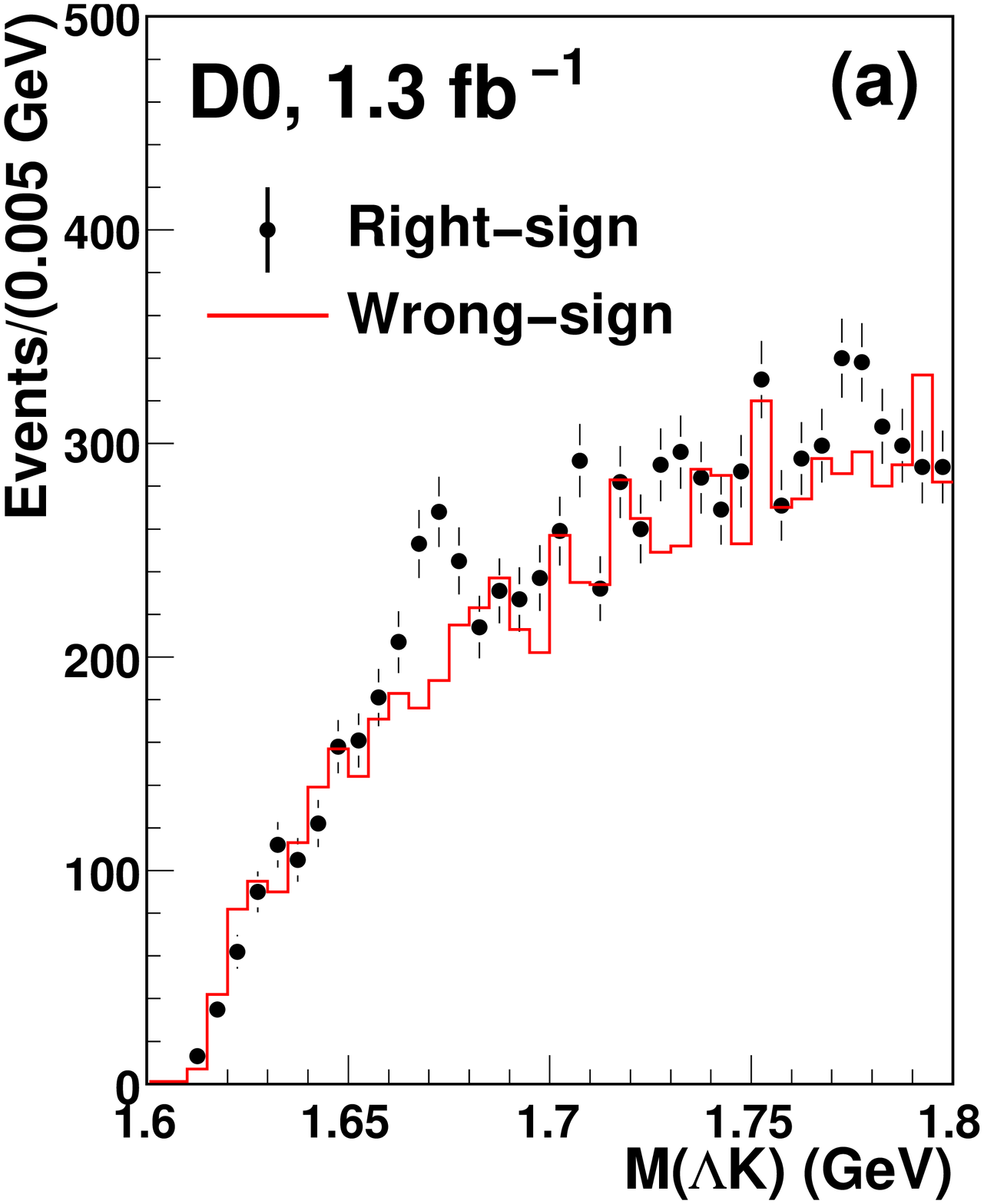,height=1.750in}
\psfig{figure=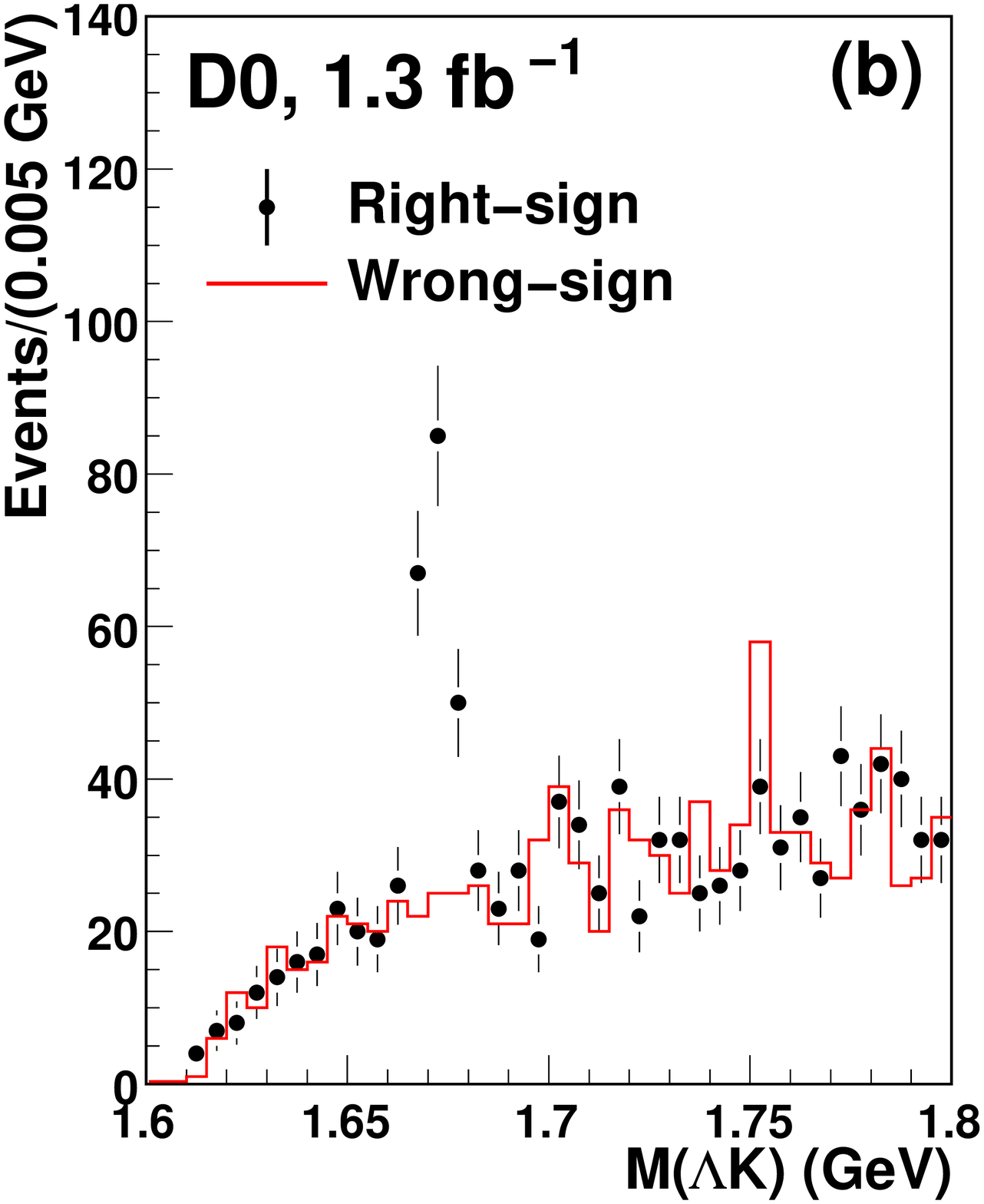,height=1.75in}
\psfig{figure=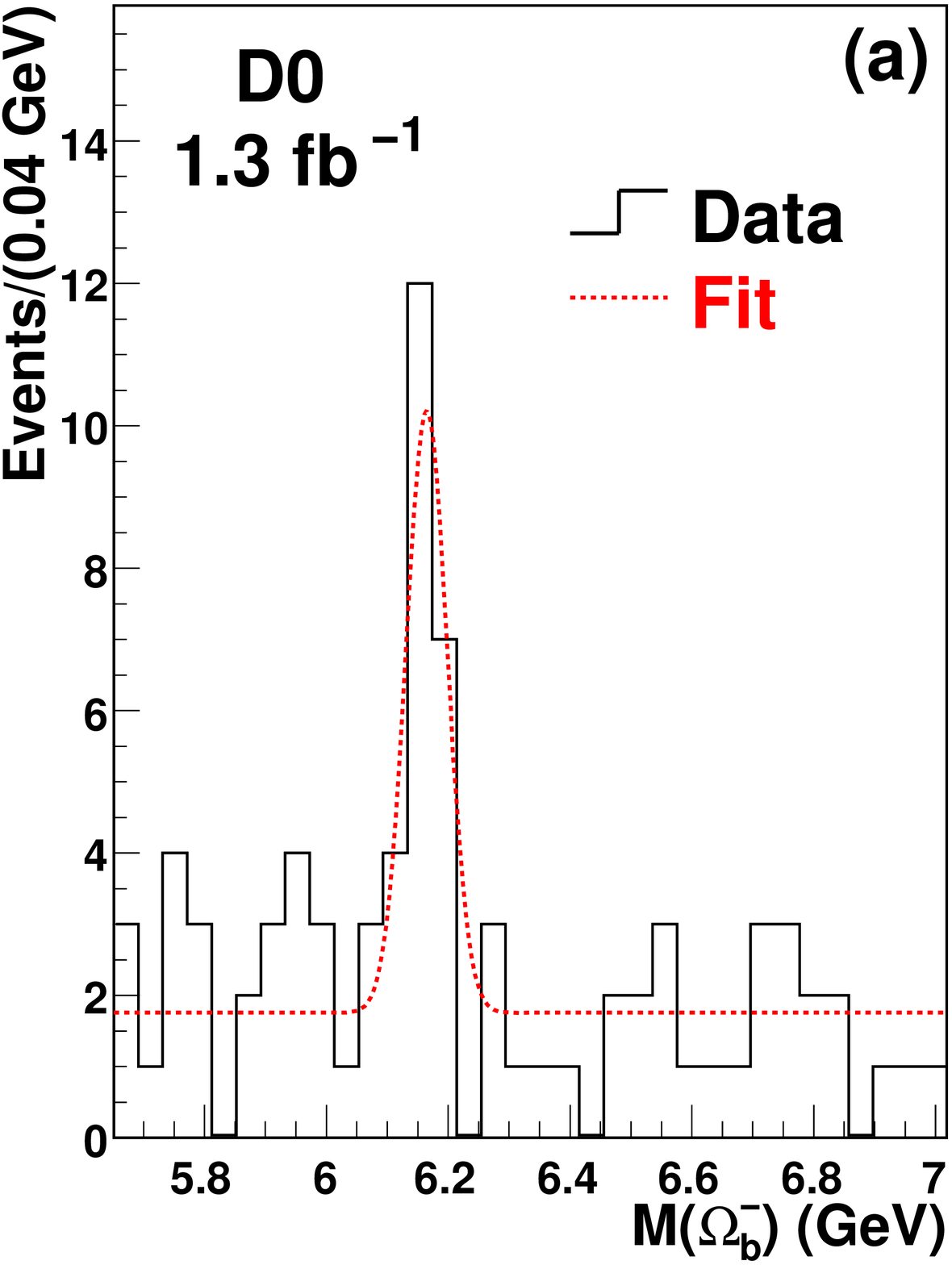,height=1.75in}
\caption{The left and center plots show $\Omega$ signal before and after the decision tree selection is applied. The right hand plot shows the mass distribution of the $\Omega_b^-$ candidates.
\label{fig:omegabmass}}
\end{center}
\end{figure}

The Tevatron has improved the understanding of B baryons through the observation of the $\Sigma^{\pm}_b$ \cite{sigmab}, $\Xi_b^{\pm}$~\cite{chib}$^{,}$\cite{chib2} and, most recently, the $\Omega_b^{-}$. These predicted baryons were not observed until Run II. D0 observes the $\Omega_b^{-}$ in the decay channel $\Omega_b^- \to J/\Psi[\to \mu^+ \mu^-]\Omega^-[\to \Lambda^0(\to p\pi^-)K^-]$ using 1.3 fb$^{-1}$ of data. To increase the acceptance of the $\Omega_b^-$ the data is reprocessed with a modified tracking algorithm that allows reconstruction of tracks with large impact parameters which occur in this decay channel, due to the long lifetime of the $\Omega$ and $\Lambda$ baryon. A boosted decision tree is used to enhance the signal of the $\Omega$ over the combinatorial background, as shown in Fig.~\ref{fig:omegabmass}. D0 observes\cite{omegab} 17.8$\pm$4.9\textit{(stat)}$\pm$0.8\textit{(syst)} $\Omega_b^-$ baryons and measures $M(\Omega_b^-)$ = 6.165$\pm$0.010\textit{(stat)}$\pm$0.013\textit{(syst)} GeV/c$^2$. The significance of the signal is greater than 5$\sigma$. The theoretical predictions of the mass~\cite{omref}, 5.94$<M(\Omega_b^-)<$6.12 GeV/c$^2$, are lower than the observed mass.
\section{B hadron lifetimes}
The lifetimes of B hadron species test the Heavy Quark Expansion theory\cite{HQE}. The predicted lifetime ratios are \cite{lifepred}$^,$\cite{lenz2}: $\tau(B^+)/\tau(B^0)$=1.053$\pm$0.016$\pm$0.017, $\tau(B_s^0)/\tau(B^0)$ = 1.00$\pm$0.01 and $\tau(\Lambda_b)/\tau(B^0)$ = 0.88$\pm$0.05. The experimental world averages for these ratios are $\tau(B^+)/\tau(B^0)$ = 1.071$\pm$0.009, $\tau(B_s^0)/\tau(B^0)$ = 0.939$\pm$0.021 and $\tau(\Lambda_b)/\tau(B^0)$ = 0.904$\pm$0.032.
\subsection{$\Lambda_b$ lifetime}
\begin{figure}
\begin{center}
\psfig{figure=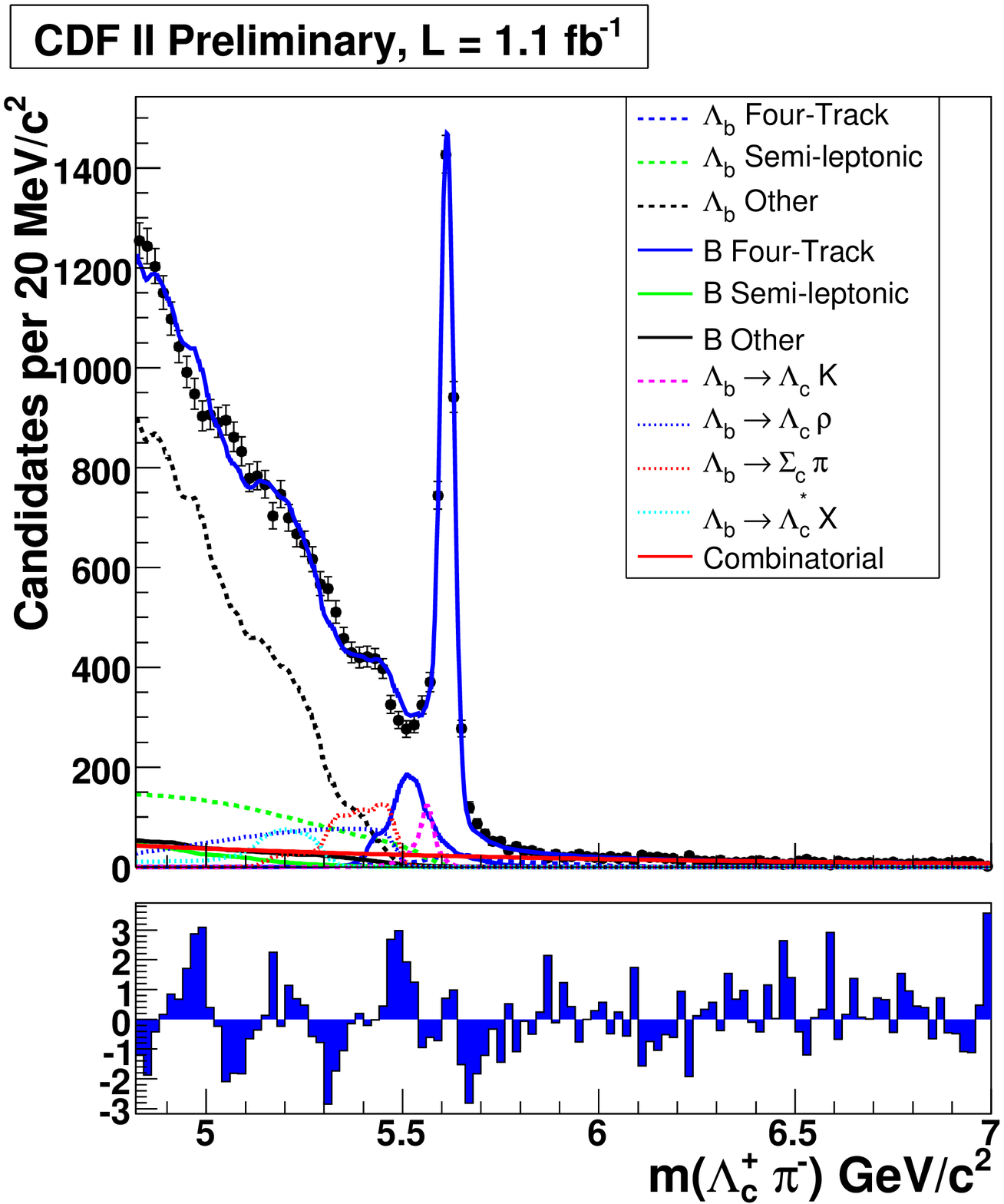,height=2.in}
\psfig{figure=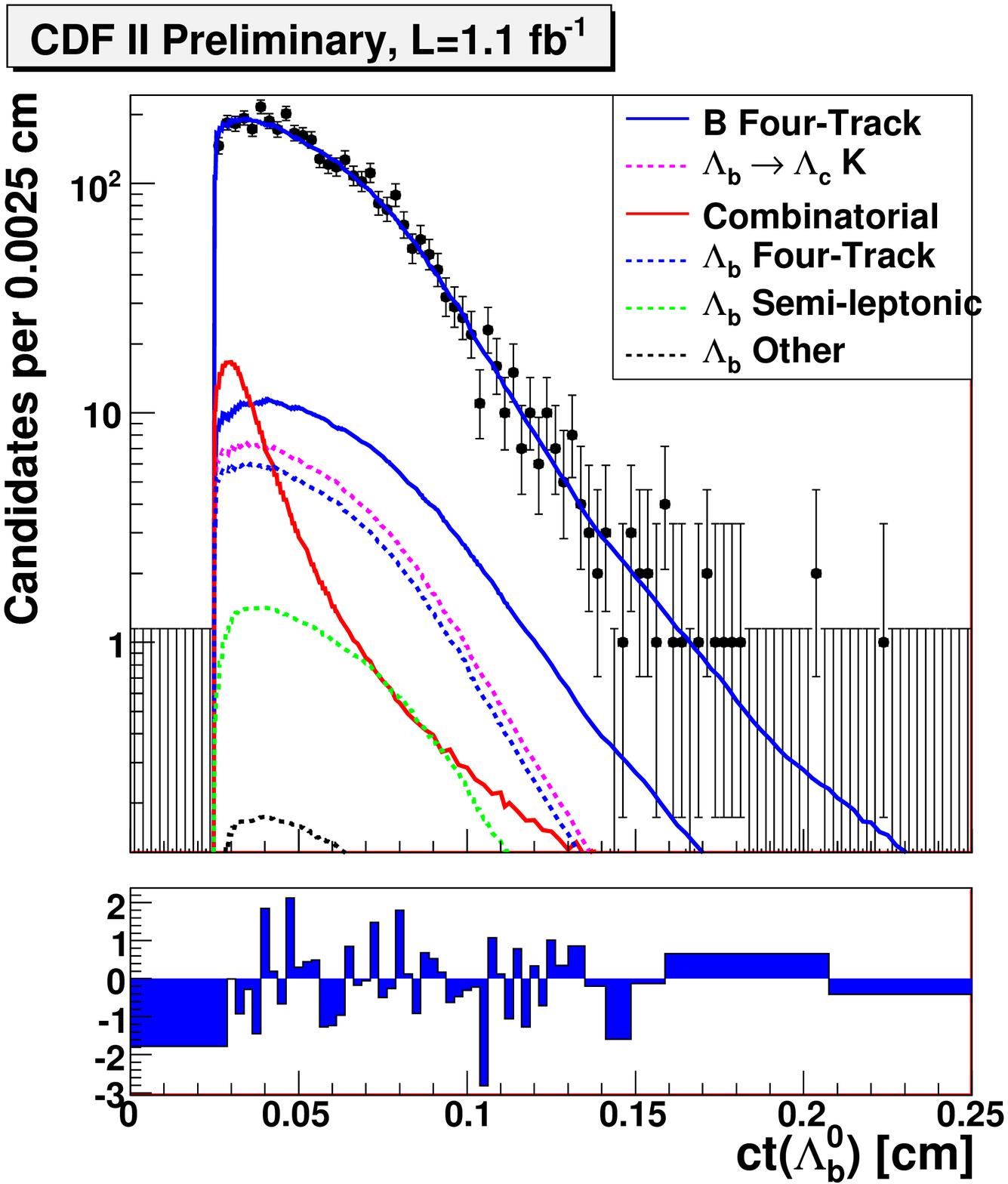,height=2.in}
\psfig{figure=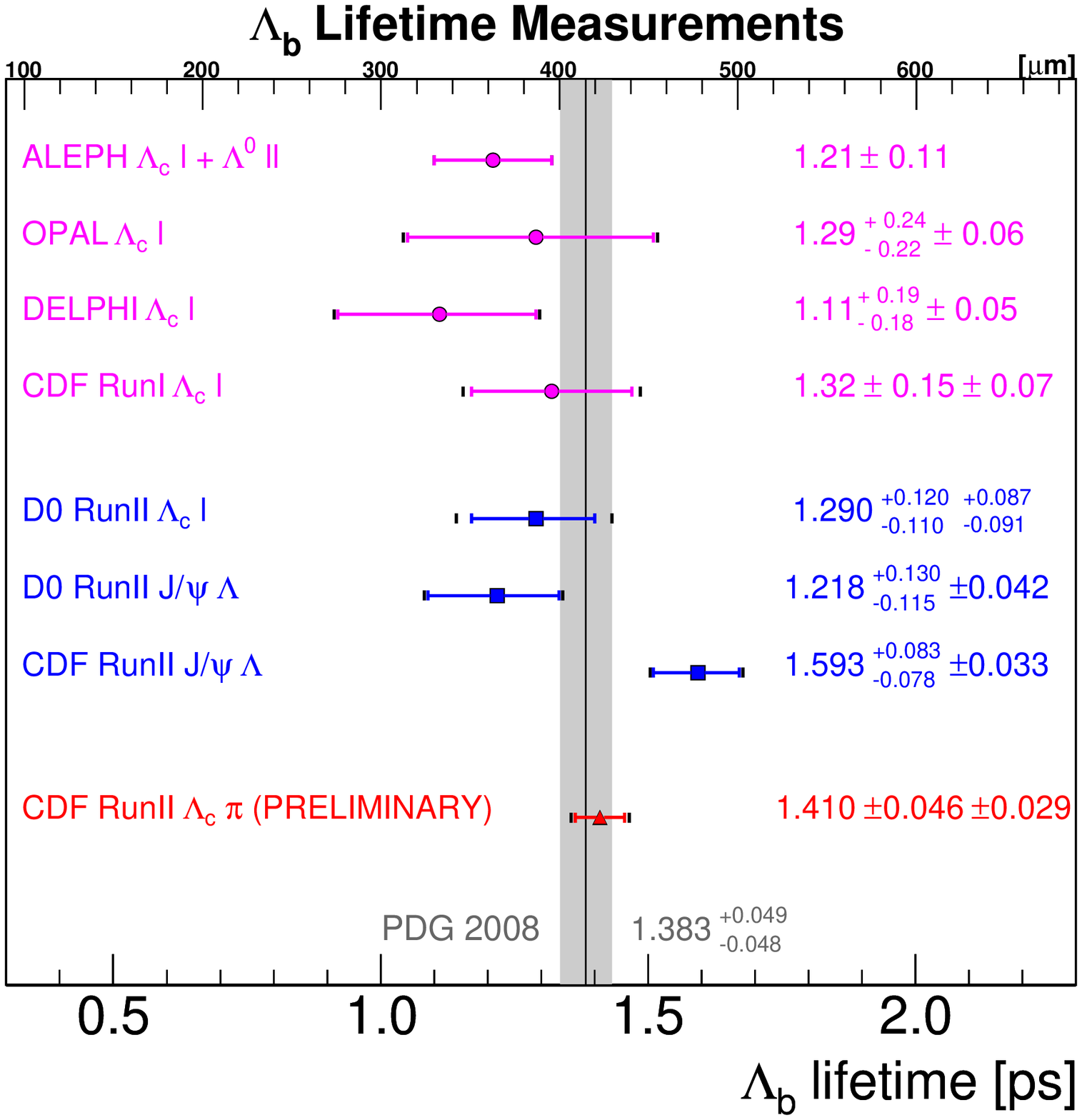,height=2.in}
\caption{The left hand plot shows the sample composition of the reconstructed $\Lambda_b$ candidates. The decay time distribution and fit projection is shown in the central plot. On the right is a comparison of $\tau(\Lambda_b)$ measurements.
\label{Lambda}}
\end{center}
\end{figure}

The $\Lambda_b$ lifetime has raised recent interest as the CDF measurement of $\tau(\Lambda_b)$ using $\Lambda_b \to J/\Psi \Lambda$~\cite{lb06} was precise but larger than all previous measurements. CDF has also measured $\tau(\Lambda_b)$ using fully reconstructed decays in the decay mode $\Lambda_b \to \Lambda_c \pi$, where the events are collected by the displaced vertex trigger\cite{lb08}. This trigger exploits the long lived nature of B hadrons and yields $\sim$3000 $\Lambda_b$ events in 1.1 fb$^{-1}$ of data after signal optimization. The sample composition is shown in Fig.~\ref{Lambda}. The trigger acceptance biases the decay time distribution. This is corrected for by modeling the trigger efficiency using Monte Carlo simulation. After the sample composition is determined, an unbinned maximum likelihood fit is performed and measures c$\tau(\Lambda_b)$= 420.1$\pm$13.7\textit{(stat)}$\pm$10.6\textit{(syst)} $\mu m$. This is the world's best determination of $\tau(\Lambda_b)$ and is consistent with all previous measurements as shown in Fig.~\ref{Lambda}. The lifetime ratio $\tau(\Lambda_b)/\tau(B^0)$ is compatible with theoretical predictions. The systematic uncertainties in this measurement are dominated by uncertainties in the simulation of the trigger response and the decay kinematics.

\subsection{$B_s^0$ lifetime}
Previous measurements of $\tau(B_s^0)$, which used semileptonic decay modes, were lower than $\tau(B^0)$. CDF has measured the $B_s^0$ flavor specific lifetime using the decay mode $B_s^0 \to D_s^- \pi^+$ using 1.3 fb$^{-1}$ of data collected by the displaced vertex trigger\cite{bslife}. As before, Monte Carlo simulation is used to correct for the decay time bias of the trigger. The measurement uses 1100 fully reconstructed decays and also 2000 partially reconstructed decays e.g. $B_s^0 \to D_s^{-*}\pi^+$, $B_s^0 \to D_s^- \rho^+$ which significantly improves the statistical precision of the result. The partially reconstructed decays have missing particles which results in a incorrect measure of the mass and transverse momentum used to determine the decay time. This is corrected for statistically by introducing a $K$ factor : $t=\frac{L_{xy}m_b^{rec}}{cP_T}\cdot K$ where the $K$ factor distribution is determined from Monte Carlo and is the ratio of the true lifetime and the lifetime measured using a partially reconstructed event. CDF measures c$\tau(B_s^0)$= 455.0$\pm$12.2\textit{(stat)}$\pm$8.2\textit{(syst)}$\mu m$ which is the most precise measurement of the flavor specific $B_s^0$ lifetime and is compatible with the world average value\cite{PDG} of $\tau(B^0)$.
\subsection{Simulation independent method of lifetime measurement}
The use of Monte Carlo to model the trigger bias leads to large systematic uncertainties that will dominate measurements with more data. A simulation free method to correct the trigger bias has been developed\cite{mcfree} at CDF to improve future measurements. This uses the decay kinematics to determine a per event efficiency. The event efficiency is evaluated, as a function of decay time, by calculating the trigger acceptance at each possible decay time along the B meson momentum direction. Using the decay mode $B^+ \to \overline{D^0} \pi^+$, we measure $\tau(B^+)$ = 498.6$\pm$6.9\textit{(stat)}$\pm$4.5\textit{(syst)} $\mu m$. This is consistent with the world average\cite{PDG}, and the measurement is presented as a proof of principle. Overall, the total systematic uncertainty is half the size of the typical error from the simulation dependent method. The largest contributions to the systematic uncertainties for the simulation free method come from assumptions on the track finding efficiencies of the trigger. 
\subsection{$B_c$ lifetime}
The lifetime of the $B_c$ meson is expected to be much smaller than the other B mesons as either the b or c quark can decay. CDF and D0 use the semileptonic decay channel $B_c \to J/\Psi l X$ (where $l=\mu,e$). The neutrino is not reconstructed and the resulting mismeasurement of the mass and transverse momentum of the $B_c$ is statistically corrected for by determining K factors from Monte Carlo. The decay time distributions of background events are modeled and calibrated using Monte Carlo and data. CDF uses both the electron and muon channels in 1.0 fb$^{-1}$ and the combined measurement\cite{CDFBC} is c$\tau(B_c)$=142$\pm$15\textit{(stat)}$\pm$6\textit{(syst)} $\mu m$. D0 analyzes the muon channel only in 1.3 fb$^{-1}$ and measures\cite{D0BC} c$\tau(B_c)$=134$\pm$11\textit{(stat)}$\pm$10\textit{(syst)} $\mu m$. Both measurements are consistent with, and more precise than, theoretical predictions\cite{bcref}.

\section{CP violation in $B_s^0$ mixing}

The study of CP violation in $B_s^0$ mixing has been carried out by both experiments in the decay $B_s^0 \to J/\Psi[\to \mu^+ \mu-] \phi[\to K^+,K^-]$. The time evolution of a mixture of the $B_s^0$ and its antiparticle $\overline{B_s^0}$ is given by the Schrodinger equation $i \frac{d}{dt} \left(\begin{smallmatrix}a\\b\end{smallmatrix}\right)=\left(M-i\frac{\Gamma}{2}\right) \left(\begin{smallmatrix}a\\b\end{smallmatrix}\right)$, where $M$ and $\Gamma$ are the $2\times 2$ mass and decay matrices that relate the flavor eigenstates, $B_s^0$ and $\overline{B_s^0}$, with the mass eigenstates, $B_s^{0H}$ and $B_s^{0L}$. The difference in mass and width between $B_s^{0H}$ and $B_s^{0L}$ is related to the off diagonal elements of the mass and decay matrices as follows: $m_s^H-m_s^L= \Delta{m_s}\approx 2|M_{12}|$, $\Gamma_L-\Gamma_H = \Delta\Gamma \approx 2|\Gamma_{12}|cos(\phi_s)$ where $\phi_s=arg(-M_{12}/\Gamma_{12})\approx 0.04$ in the Standard Model(SM)\cite{lenz1}. Any new particle participating in the mixing loop that carries a weak phase will be measurable given that the SM prediction is $\sim$0. This phase $\phi^{NP}$ is accessible through the measurement of $\beta_s^{J/\Psi\phi}$ which is the relative phase between the direct decay amplitude and mixing followed by decay amplitude. In the SM $\beta_s = arg \left(\frac{-V_{ts}V_{tb}^*}{V_{cs}V_{cb}^*}\right) \approx 0.02$ where $V_{ij}$ are the elements of the CKM matrix. A large observed $\beta_s$ is a clear indication of new physics, and is related to the new physics phase, $\phi^{NP}$, through $2\beta_s^{J/\Psi\phi}=2\beta_s^{SM} - \phi_s^{NP}$.

\begin{figure}
\begin{center}
\psfig{figure=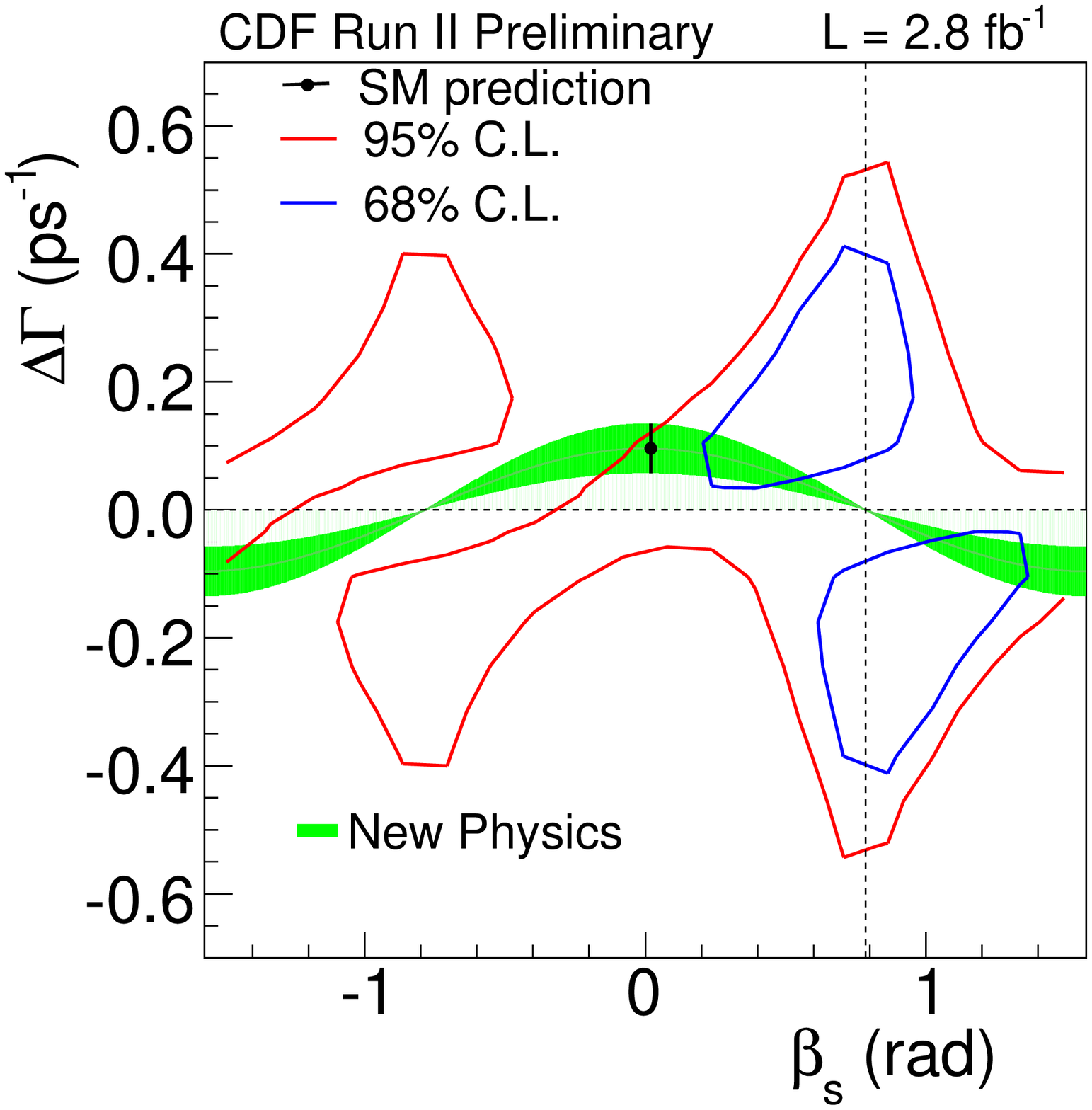,height=1.75in}
\psfig{figure=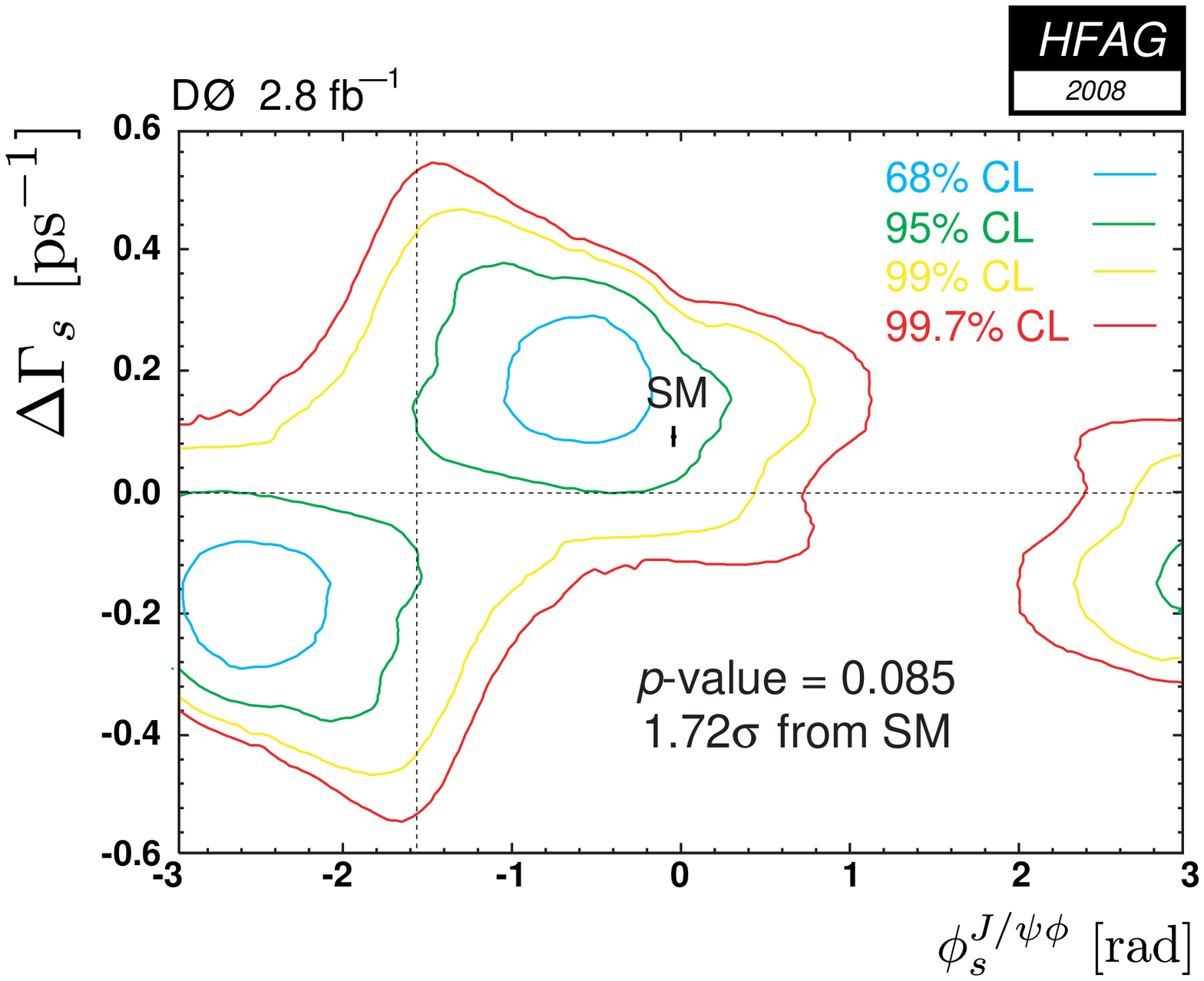,height=1.75in}
\caption{Confidence intervals on the $\Delta \Gamma - \beta_s$ plane. Note that $\phi_s=-2\beta_s$ and hence the shift with respect to the SM expectation is in the same direction for each experiment.
\label{fig:radish}}
\end{center}
\end{figure}

CDF and D0 reconstruct 3200 and 2000 signal events in 2.8 fb$^{-1}$ of data respectively. The angular distribution of the muons and kaons, (to disentangle the CP of final state), the tagging information, (to determine the initial flavor of the meson), mass, decay time and mixing frequency are put together in an unbinned maximum likelihood fit to extract $\beta_s$ and $\Delta\Gamma$. The results are presented in confidence intervals as the errors are non Gaussian in Fig.~\ref{fig:radish}. CDF observes a 1.8$\sigma$ deviation\cite{cpvcdf} from the expected SM values and D0 observes a 1.7$\sigma$ deviation\cite{cpvd0}. A combination of the D0 result and a previous CDF result\cite{cdfold} using only 1.35 fb$^{-1}$ of data gives a 2.2$\sigma$ deviation from the SM. Although not statistically significant, it is interesting that both experiments observe similar shifts. The evolution of this measurement as more data and analysis improvements are incorporated is awaited due to the potential to observe new physics.  

\section{Conclusions}
The Tevatron continues to lead the way with B-baryon mass measurements and B hadron lifetime measurements. New and improved results can be expected with the data now available. CDF and D0 have made interesting measurements related to CP violation in $B_s^0$ mixing and could potentially observe the signature of new physics in the future. These exciting results demonstrate the success of the B physics program at the Tevatron.

\end{document}